\documentclass[conference]{IEEEtran}
\usepackage{amsmath}
\usepackage{cite}
\usepackage{graphicx}
\usepackage{epstopdf}
\usepackage{amsfonts,amsmath,amssymb}
\usepackage{graphicx}
\usepackage{bm,bbm}
\usepackage{subfigure}
\usepackage{stfloats}
\usepackage{color}
\usepackage{algorithmic}
\usepackage{algorithm}

\newtheorem{theorem}{\textbf{Theorem}}

\newtheorem{proposition}{\textbf{Proposition}}

\begin{document}

\title{On Channel Reciprocity to Activate Uplink Channel Training for Downlink Wireless Transmission in Tactile Internet Applications}

\author{\IEEEauthorblockN{Chunhui Li, Shihao Yan, and Nan Yang}
\IEEEauthorblockA{Research School of Engineering, The Australian National University, Canberra, ACT 2601, Australia}
\IEEEauthorblockA{Emails: \{chunhui.li, shihao.yan, nan.yang\}@anu.edu.au}}

\markboth{Submitted to IEEE ICC 2018 Workshop}{Li \MakeLowercase{\textit{et
al.}}: On Channel Reciprocity to Activate Uplink Channel Training for Downlink Wireless Transmission in Tactile Internet Applications}

\maketitle

\begin{abstract}
We determine, for the first time, the requirement on channel reciprocity to activate uplink channel training, instead of downlink channel training, to achieve a higher data rate for the downlink transmission from a multi-antenna base station to a single-antenna user. We first derive novel closed-form expressions for the lower bounds on the data rates achieved by the two channel training strategies by considering the impact of finite blocklength. The performance comparison result of these two strategies is determined by the amount of channel reciprocity that is utilized in the uplink channel training. We then derive an approximated expression for the minimum channel reciprocity that enables the uplink channel training to outperform the downlink channel training. Through numerical results, we demonstrate that this minimum channel reciprocity decreases as the blocklength decreases or the number of transmit antennas increases, which shows the necessity and benefits of activating the uplink channel training for short-packet communications with multiple transmit antennas. This work provides pivotal and unprecedented guidelines on choosing channel training strategies and channel reciprocity calibrations, offering valuable insights into latency reduction in the Tactile Internet applications.
\end{abstract}

\section{Introduction}\label{introduction}

The fifth generation (5G) wireless communications is envisioned to establish a fully connected wireless society, providing enhanced mobile broadband (eMBB) service, enabling massive machine-type communications (mMTC), and supporting mission-critical communications (MCC)~\cite{Andrews2014,Dahlman2014,Chen2014}.
Specifically, the mMTC will provide ubiquitous connectivity for an enormous amount of low-cost devices and the MCC will enable real-time data transmission with stringent requirements on latency and reliability. Due to its unique capabilities, 5G will play a dominant role in supporting the Tactile Internet. The Tactile Internet is a network which combines ultra-low latency with extremely high availability, reliability, and security to enable real-time interaction between humans and machines via tactile sensations, thus being a key driver for economic innovation and society development~\cite{Fettweis2014,Simsek2016}.

As a highly promising technology to achieve low latency in supporting Tactile Internet applications, short-packet communications (SPC) has recently attracted growing research attention~\cite{Popovski2014,Durisi2016,Durisi2016TOC}. Short packets are the typical forms of the traffic generated by sensors and small mobile devices involved in MTC. For example, in industrial manufacturing and control systems, measurements and control commands are of small size (e.g., 10 to 20 bytes)~\cite{Johansson2015,Yilmaz2015,Ashraf2016} and need to be communicated with ultra-low latency (e.g., 0.5 to 1 ms) and ultra-high reliability (e.g., the decoding error probability is on the order of $10^{-9}$). The current frameworks used to analyze wireless communications systems cannot be directly adopted to examine SPC, since these frameworks assume that the size of metadata (e.g., the data used for channel training) is negligible comparing to the large packet size~\cite{Shannon1948}.

In order to achieve ultra-low latency, the overhead in SPC has to be redesigned, since this overhead is not negligible and may become the dominating factor for latency. Within the overhead, the part used for channel training which enables the transceivers to learn the channel state information (CSI) is unavoidable and could dominate the total overhead. For the time division duplex (TDD) wireless communications from a base station (BS) to a user, there are two channel training strategies before data transmission. In the first downlink channel training strategy, the BS transmits pilot signals to enable the user for estimating the downlink CSI from the BS to the user and then the user feeds back the estimation to the BS. In the second uplink channel training strategy, the user transmits pilot signals to enable the BS for estimating the uplink CSI from the user to the BS and then the BS learns the downlink CSI based on the channel reciprocity between the uplink and the downlink.

Although the aforementioned two strategies are widely adopted in wireless communications, their performance has never been examined in the context of SPC. The most pressing challenge here is that the impact of the finite blocklength needs to be considered, where the achievable data rates have never been derived by considering the cost of different channel training strategies and the effect of the channel dispersion. It is well known that the benefits of utilizing channel reciprocity to enable uplink channel training in multiple-input multiple-output (MIMO) systems scale linearly with the number of transmit antennas at the BS~\cite{Jose2011,Marzetta2010}. This is due to the fact that the resources (e.g., time slots) used in the downlink channel training are linear functions of the number of transmit antennas~\cite{Jose2011}, while the ones used in the uplink channel training are independent of this number~\cite{Marzetta2010}. However, achieving channel reciprocity in practical scenarios requires appropriate hardware calibrations to compensate for the unknown amplitude scaling and phase shift between the downlink and uplink channels~\cite{Kaltenberger2010}. Meanwhile, we note that the performance of uplink channel training highly depends on the amount of the achieved channel reciprocity. Against this background, a never-before-answered question is \emph{``How much channel reciprocity is required to guarantee the uplink channel training to outperform the downlink channel training?''}, which motivates this work. The answer to this question is pivotal for reducing the wireless transmission latency in the Tactile Internet applications.

In order to fully tackle this question, in this work we first examine the performance of the downlink and uplink channel training strategies. Specifically, we derive, \emph{for the first time}, closed-form expressions for the lower bounds on the data rates achieved by these two strategies. These expressions allow us to determine the minimum channel reciprocity that is required to ensure a higher data rate achieved by the uplink channel training relative to the downlink channel training. Specifically, we determine an analytical expression to approximate this minimum channel reciprocity to draw useful insights into the affecting parameters. Our examination first indicates that this minimum channel reciprocity decreases as the total blocklength decreases, which demonstrates the superiority of the uplink channel training in the context of SPC. As expected, our results show that this minimum channel reciprocity also decreases as the number of transmit antennas at the BS increases. This indicates that the uplink channel training becomes more desirable and easier to achieve when a large number of transmit antennas are deployed at the BS. The derived minimum channel reciprocity provides practical guidelines on choosing channel training strategies and channel reciprocity calibrations.

\section{System Model}\label{sec:system_model}

We consider a multiple input single output (MISO) communications system where an $N_{\mathrm{B}}$-antenna BS communicates with a single-antenna user. We denote $\mathbf{h_{\mathrm{u}}}$ as the $N_\mathrm{B} \times 1$ uplink channel vector from the user to the BS and denote $\mathbf{h}_{\mathrm{d}}$ as the $1 \times N_\mathrm{B}$ downlink channel vector from the BS to the user. All the channels are subject to independent quasi-static Rayleigh fading with the finite blocklength $T$. We assume that the channels remain constant during one fading block. The entries of $\mathbf{h_{\mathrm{u}}}$ and $\mathbf{h}_{\mathrm{d}}$ are assumed to be independent and identically distributed (i.i.d.) circularly symmetric complex Gaussian random variables with zero mean and unit variance, i.e., $\mathbf{h}_{\mathrm{d}}\sim\mathcal{CN} \left(0,\mathbf{I}_{N_{\mathrm{B}}}\right)$ and $\mathbf{h}_{\mathrm{u}}\sim\mathcal{CN} \left(0,\mathbf{I}_{N_{\mathrm{B}}}\right)$, where $\mathcal{CN}\left(\mu,\nu\right)$ is the complex Gaussian distribution with the mean of $\mu$ and the variance of $\nu$ and $\mathbf{I}_{N_{\mathrm{B}}}$ is an $N_{\mathrm{B}} \times N_{\mathrm{B}}$ identity matrix. Furthermore, we assume that the transmit power $P_{\mathrm{b}}$ at the BS is fixed for each channel use. Additionally, we assume that the user and the BS have the knowledge about the statistical information of all the channels.

\subsection{Channel Training}

\subsubsection{Uplink Channel Training}

In the uplink channel training, $\mathbf{h}_{\mathrm{u}}$ is obtained at the BS via uplink training where the user sends pilot sequences to the BS for estimating $\mathbf{h}_{\mathrm{u}}$. Considering the channel reciprocity between the uplink and the downlink, the downlink channel vector can be expressed as a function of the uplink channel vector, given by \cite{Nosrat-Makouei2011,Mi2017}
\begin{align}\label{h_d_open-loop}
\mathbf{h}_{\mathrm{d}}=\sqrt{\phi}\mathbf{h}_{\mathrm{u}}^{T}+\sqrt{1-\phi}\mathbf{e}^{T},
\end{align}
where $\phi$ is defined as the channel reciprocity coefficient between the uplink and the downlink, $\mathbf{h}_{\mathrm{u}}^{T}$ represents the transpose of $\mathbf{h}_{\mathrm{u}}$, and $\mathbf{e}$ is the $N_\mathrm{B} \times 1$ vector which reflects the uncertain part of $\mathbf{h}_{\mathrm{u}}$. The entries of $\mathbf{e}$ are i.i.d. and each of them follows $\mathcal{CN}\left(0,1\right)$.

We note that the value of $\phi$ quantifies the level of channel reciprocity, where $0\leq\phi\leq1$. In practical scenarios, the level of channel reciprocity is determined by the uplink channel estimation error and the frequency offset between the transmitter and receiver~\cite{Kaltenberger2010}. Specifically, $\phi=1$ indicates that the perfect channel reciprocity is achieved such that the downlink channel is exactly the same as the uplink channel. When $\phi$ decreases, the channel reciprocity becomes less reliable. When $\phi=0$, the channel reciprocity does not exist such that the downlink channel is independent of the uplink channel.

\subsubsection{Downlink Channel Training}

In the downlink channel training, $\mathbf{h}_{\mathrm{d}}$ is obtained at the BS via downlink training and uplink feedback where the BS sends pilot sequences to the user for estimating $\mathbf{h}_{\mathrm{d}}$ and then the user feeds back the estimate. When the BS sends pilot sequences in $T_{\mathrm{tr}}$ symbol periods, the received signal at the user is given by 
\begin{align}\label{y_d_training}
\mathbf{y}_{\mathrm{d}}=\sqrt{\Lambda}
\mathbf{h}_{\mathrm{d}}\mathbf{S}_{\mathrm{d}}+\mathbf{n}_{\mathrm{d}},
\end{align}
where $\Lambda\triangleq{T_{\mathrm{tr}}P_{\mathrm{b}}}/{N_\mathrm{B}}$, $\mathbf{y}_{\mathrm{d}}$ is the $1 \times T_{\mathrm{tr}}$ received signal vector, $\mathbf{S}_{\mathrm{d}}$ is the $N_{\mathrm{B}} \times T_{\mathrm{tr}}$ pilot sequence matrix transmitted by the BS which satisfies $\mathbf{S}_{\mathrm{d}}\mathbf{S}_{\mathrm{d}}^{H}=\mathbf{I}_{N_\mathrm{B}}$, and $\mathbf{n_{\mathrm{d}}}$ is the $1 \times T_{\mathrm{tr}}$ additive white Gaussian noise (AWGN) vector at the user with i.i.d entries following $\mathcal{CN}\left(0,\sigma_{\mathrm{u}}^2\right)$. We assume that the linear minimum mean square error (MMSE) estimator is adopted at the user. Based on the known pilot sequences, the user obtains the estimates of $\mathbf{h_{\mathrm{d}}}$ as \cite{Hassibi2003,Yan2017}
\begin{align}\label{h_hat_down}
\mathbf{\hat{h}}_{\mathrm{d}}&=\frac{\sqrt{\Lambda}}{\Lambda+\sigma_{\mathrm{u}}^2} \mathbf{y}_{\mathrm{d}}\mathbf{S}^{H}_{\mathrm{d}}.
\end{align}
As per the rules of the linear MMSE estimator, the entries of $\mathbf{\hat{h}}_{\mathrm{d}}$ are i.i.d. and each of them follows $\mathcal{CN}\left(0,\sigma_{\mathrm{\hat{h}}_{\mathrm{d}}}^2\right)$, where $\sigma_{\mathrm{\hat{h}}_{\mathrm{d}}}^{2}=\Lambda/\left(\Lambda+\sigma_{\mathrm{u}}^2\right)$.
We note that the estimation error, given by $\mathbf{\hat{e}}_{\mathrm{d}}=\mathbf{h}_{\mathrm{d}}-\mathbf{\hat{h}}_{\mathrm{d}}$, is independent of the estimate $\mathbf{\hat{h}_{\mathrm{d}}}$. We also note that the entries of $\mathbf{\hat{e}_{\mathrm{d}}}$ are i.i.d. and each follows $\mathcal{CN}\left(0,\sigma_{\mathrm{\hat{e}}_{\mathrm{d}}}^{2}\right)$, where $\sigma_{\mathrm{\hat{e}}_{\mathrm{d}}}^{2}=\sigma_{\mathrm{u}}^2/\left(\Lambda+\sigma_{\mathrm{u}}^2\right)$.
We note that $T_{\mathrm{tr}}\geq{}N_\mathrm{B}$ needs to be ensured in the system, in order to obtain a reliable estimate of $\mathbf{h}_{\mathrm{d}}$.

\subsection{Data Transmission}

After obtaining the downlink channel vector through either the uplink channel training or the downlink channel training, the BS selects an $N_\mathrm{B} \times1 $ normalized beamforming vector $\mathbf{v}$ to transmit signals to the user. The BS uses the obtained CSI as it is perfect. Therefore, the transmitted signal $\mathbf{x}$ is written as $\mathbf{x}=\mathbf{v}u$, where $u$ is the information signal transmitted from the BS to the user. The received signal at the user in one symbol period is given by $y=\sqrt{P_{\mathrm{b}}}\mathbf{h}_\mathrm{d}\mathbf{v}u+n$,
where $n$ is the AWGN at the user with zero mean and variance $\sigma_{\mathrm{u}}^2$, while $\mathbf{x}$ is subject to the average power constraint $\mathbb{E}\left[\|\mathbf{x}\|^{2}\right]=1$ with $\mathbb{E}\left[\cdot\right]$ denoting expectation.

\subsection{Data Rate with Finite Blocklength}

Considering finite-blocklength transmission, the achievable data rate in the fading channel can be tightly approximated as $R\left(T,\epsilon,\gamma\right)$ which is a function of the blocklength (i.e., the number of channel use) $T$, the decoding error probability $\epsilon$, and the signal-to-noise ratio (SNR) $\gamma$~\cite{Durisi2016TOC}. Mathematically, this function is given by~\cite{Polyanskiy2010}
\begin{align}\label{The maximum coding rate}
R\left(T,\epsilon,\gamma\right)\approx
C\left(\gamma\right)-\sqrt{\frac{1}{T}V\left(\gamma\right)}Q^{-1}\left(\epsilon\right),
\end{align}
where $C\left(\gamma\right)=\log_{2}\left(1+\gamma\right)$ is the channel capacity, $V\left(\gamma\right)=\left(\log_{2}e\right)^{2}\left(1-\left(1+\gamma\right)^{-2}\right)$ is the channel dispersion, and $Q^{-1}\left(\cdot\right)$ is the inverse $Q$-function. We note that \eqref{The maximum coding rate} is tight, even for a relatively small $T$, e.g., $T=100$~\cite{Polyanskiy2010}.

\section{Achievable Data Rates of Uplink and Downlink Channel Training Strategies}\label{sec:perfect scheme}

In this section, we focus on an ideal scenario, where the uplink channel training is perfect (i.e., no channel estimation error) and does not cost any time slot, meanwhile the feedback in the downlink channel training is perfect (i.e., no feedback error) and of no cost in terms of time slots. We would like to clarify that in this scenario, the ignored cost of time slots in two channel training strategies may not be the same. Notably, in general this cost is higher for the downlink channel training than that for the uplink channel training. In the uplink channel training, the user can use only one time slot to send pilot sequences and the BS only has to feed $||\mathbf{h}||$, but not $\mathbf{h}$, back to the user. In the downlink channel training, however, the user has to feed  $\mathbf{h}$ (i.e., $N_{\textrm{B}}$ complex numbers) back to the BS. We next derive lower bounds on the data rates achieved by uplink channel training and downlink channel training.


\subsection{Achievable Data Rate of Uplink Channel Training}

Under the assumption of perfect uplink channel training, the BS has the perfect knowledge about the uplink channel. As such, the beamforming vector is selected as $\mathbf{v}=\mathbf{h}_{\mathrm{u}}/\|\mathbf{h}_{\mathrm{u}}\|$ and the received signal at the user in one symbol period is given by
\begin{align}\label{y_op}
y&=\sqrt{P_{\mathrm{b}}}\mathbf{h_{\mathrm{d}}}\mathbf{x}+n
=\sqrt{P_{\mathrm{b}}}\left(\sqrt{\phi}~\mathbf{h}_{\mathrm{u}}^{T} + \sqrt{1-\phi}~\mathbf{e}^{T} \right)\mathbf{v}u+n\notag\\
&=\sqrt{P_{\mathrm{b}}\phi}\mathbf{h}_{\mathrm{u}}^{T}\frac{\mathbf{h}_{\mathrm{u}}}{\left\|\mathbf{h}_{\mathrm{u}}\right\|}u +\underbrace{\sqrt{P_{\mathrm{b}}\left(1-\phi\right)}\mathbf{e}^{T} \frac{\mathbf{h}_{\mathrm{u}}}{\left\|\mathbf{h}_{\mathrm{u}}\right\|}u+n}_{\tilde{n}}.
\end{align}
Following \cite{Hassibi2003}, in this work we consider the worst-case scenario for decoding at the user, where $\tilde{n}$ in \eqref{y_op} is approximated as the zero-mean Gaussian noise. Under this approximation, the achievable data rate derived below is a lower bound. As per \eqref{y_op}, the signal-to-interference-plus-noise ratio (SINR) at the user is given by
\begin{align}\label{SINR_perfect uplink channel estimation}
\gamma_{\mathrm{u}} &=\frac{P_{\mathrm{b}}\phi {\left\|\mathbf{h}_{\mathrm{u}}\right\|}^2}{P_{\mathrm{b}} \left(1-\phi\right)\frac{{\left|\mathbf{e}^{T} \mathbf{h}_{\mathrm{u}} \right|}^2}{{\left\|\mathbf{h}_{\mathrm{u}}\right\|}^2} + \sigma_{\mathrm{u}}^2}
=\frac{P_{\mathrm{b}} \phi \left\|\mathbf{h_{\mathrm{u}}} \right\|^{2}}{\sigma_{\mathrm{\tilde{n}}}^2},
\end{align}
where $\sigma_{\mathrm{\tilde{n}}}^2$ is the variance of $\tilde{n}$, given by $\sigma_{\mathrm{\tilde{n}}}^2=P_{\mathrm{b}}\left(1-\phi\right)+\sigma_{\mathrm{u}}^2$.
Considering finite-blocklength transmission, for a given $\epsilon$ the lower bound on the data rate can be approximated by \cite{Polyanskiy2010}
\begin{align}\label{R_u}
R_{\mathrm{u}}&= \mathbb{E}_{\left\|\mathbf{h_{\mathrm{u}}} \right\|^{2}}\left[C\left(\gamma_{\mathrm{u}}\right)
-\sqrt{\frac{1}{T} V\left(\gamma_{\mathrm{u}}\right)}~Q^{-1}(\epsilon)\right]\notag\\
&=\mathbb{E}_{\left\|\mathbf{\bar{h}_{\mathrm{u}}} \right\|^{2}}\left[C\left(\gamma^{\mathrm{u}}_{\mathrm{eff}}\left\|\mathbf{\bar{h}_{\mathrm{u}}} \right\|^{2}\right)\right]\notag\\
&~~~-\mathbb{E}_{\left\|\mathbf{\bar{h}_{\mathrm{u}}} \right\|^{2}}\left[\sqrt{\frac{1}{T} V\left(\gamma^{\mathrm{u}}_{\mathrm{eff}} \left\|\mathbf{\bar{h}_{\mathrm{u}}} \right\|^{2}\right)}~Q^{-1}\left(\epsilon\right)\right],
\end{align}
where $\mathbf{\bar{h}_{\mathrm{u}}}\triangleq\mathbf{h_{\mathrm{u}}}/\sigma_{\mathrm{h}_{\mathrm{u}}}$ is the normalized channel vector, $\sigma_{\mathrm{h}_{\mathrm{u}}}$ is the standard deviation of $\mathbf{h_{\mathrm{u}}}$, $\mathbf{\bar{h}_{\mathrm{u}}}\sim\mathcal{CN}\left(0,\mathbf{I}_{N_{\mathrm{B}}}\right)$, $\gamma^{\mathrm{u}}_{\mathrm{eff}}=\frac{\rho_{\mathrm{b}} \phi } {\left(1-\phi\right) \rho_{\mathrm{b}} +1}$ is the effective SNR, and $\rho_{\mathrm{b}} ={P_{\mathrm{b}}}/{\sigma_{\mathrm{u}}^2}$ is the average SNR.

In the following theorem, we derive a closed-form expression for the lower bound on the data rate achieved by the uplink channel training.
\begin{theorem}\label{Theorem 1_Description}
The lower bound on the data rate achieved by the uplink channel training is derived as
\begin{align}\label{Theorem 1}
R_{\mathrm{u}}
=\Phi\left(\gamma^{\mathrm{u}}_{\mathrm{eff}},N_\mathrm{B}\right)
-\Psi\left(\gamma^{\mathrm{u}}_{\mathrm{eff}},N_\mathrm{B},T\right),
\end{align}
where the functions $\Phi\left(\gamma^{\mathrm{u}}_{\mathrm{eff}},N_\mathrm{B}\right)$ and $\Psi\left(\gamma^{\mathrm{u}}_{\mathrm{eff}},N_\mathrm{B},T\right)$ are given by \eqref{Phi} and \eqref{Psi}, respectively, on the top of next page.
\begin{IEEEproof}
The proof is presented in Appendix~\ref{Theorem 1 Proof}.
\end{IEEEproof}
\end{theorem}

\emph{\textbf{Theorem}~\ref{Theorem 1_Description}} presents a channel-independent and accurate expression for the lower bound on the data rate achieved by the uplink channel training, as will be shown in Section \ref{sec:numerical results}. This expression allows us to compare the performance of the uplink channel training and downlink channel training efficiently.

\begin{figure*}
\begin{align}\label{Phi}
\Phi\left(\gamma_{\mathrm{eff}},N_\mathrm{B}\right)=\frac{e^{\frac{1}{\gamma_{\mathrm{eff}}}}}
{\ln{2}~\Gamma(N_\mathrm{B})~\gamma_{\mathrm{eff}}^{N_\mathrm{B}}}
\sum_{i=0}^{N_\mathrm{B}-1} \binom{N_\mathrm{B}-1}{i} (-1)^{N_\mathrm{B}-1-i}~\textbf{G}_{2,3}^{3,0}\left({\begin{array}{c} -i,-i \\ 0,-1-i,-1-i \end{array}} \;\middle|\; \frac{1}{\gamma_{\mathrm{eff}}} \right).
\end{align}\vspace{-0.5cm}
\end{figure*}

\begin{figure*}
\begin{align}\label{Psi}
\Psi\left(\gamma_{\mathrm{eff}},N_\mathrm{B},T\right)
=\sqrt{\frac{2\pi}{T}}~\frac{Q^{-1}(\epsilon)}{\Gamma(N_\mathrm{B})~\ln{2}}~
e^{-(N_{\mathrm{B}}-1)}~(N_{\mathrm{B}}-1)^{N_{\mathrm{B}}-\frac{1}{2}}~
\sqrt{1-\left(1+\gamma_{\mathrm{eff}}\left(N_{\mathrm{B}}-1\right)\right)^{-2}}.
\end{align}
\hrulefill
\end{figure*}

\subsection{Achievable Data Rate of Downlink Channel Training}

As assumed in the ideal scenario, the feedback from the user to the BS is perfect. As such, the BS has $\mathbf{\hat{h}}_{\mathrm{d}}$ and the beamforming vector is selected as $\mathbf{v}=\mathbf{\hat{h}}_{\mathrm{d}} /\|\mathbf{\hat{h}_{\mathrm{d}}}\|$. Then, the received signal at the user in one symbol period is given by
\begin{align}\label{y_cl}
y&=\sqrt{P_{\mathrm{b}}}\mathbf{h}_{\mathrm{d}}\mathbf{x}+n
=\sqrt{P_{\mathrm{b}}}\left(\mathbf{\hat{h}}_{\mathrm{d}}+\mathbf{\hat{e}}_{\mathrm{d}}\right)\mathbf{v}u+n\notag\\
&=\sqrt{P_{\mathrm{b}}}\mathbf{\hat{h}}_{\mathrm{d}} \frac{\mathbf{\hat{h}}_{\mathrm{d}}}{\left\|\mathbf{\hat{h}_{\mathrm{d}}}\right\|}u +\underbrace{\sqrt{P_{\mathrm{b}}}\mathbf{\hat{e}}_{\mathrm{d}} \frac{\mathbf{\hat{h}}_{\mathrm{d}}}{\left\|\mathbf{\hat{h}_{\mathrm{d}}}\right\|}u+n}_{\hat{n}}.
\end{align}
Once again, we consider the worst-case scenario for decoding at the user where $\hat{n}$ in \eqref{y_cl} is approximated as Gaussian. Accordingly, the SINR at the user is given by
\begin{align}\label{SINR_perfect CSI}
\gamma_{\mathrm{d}} &=\frac{P_{\mathrm{b}}{\left\|\mathbf{\hat{h}}_{\mathrm{d}} \right\|}^2}{P_{\mathrm{b}} \frac{{\left|\mathbf{\hat{e}}_{\mathrm{d}}\mathbf{\hat{h}}_{\mathrm{d}} \right|}^2}{{\left\|\mathbf{\hat{h}}_{\mathrm{d}} \right\|}^2} + \sigma_{\mathrm{u}}^2}=\frac{P_{\mathrm{b}} \|\mathbf{\hat{h}_{\mathrm{d}}}\|^{2}}{\sigma_{\mathrm{\hat{n}}}^2},
\end{align}
where $\sigma_{\mathrm{\hat{n}}}^2$ is the variance of $\hat{n}$, given by $\sigma_{\mathrm{\hat{n}}}^2=P_{\mathrm{b}}\sigma_{\mathrm{\hat{e}_{\mathrm{d}}}}^2+\sigma_{\mathrm{u}}^2$.
Then, the lower bound on the data rate achieved by the downlink channel training is written as
\begin{align}\label{R_d}
R_{\mathrm{d}}&=\left(1-\frac{T_{\mathrm{tr}}}{T}\right)
\mathbb{E}_{\left\|\mathbf{\hat{h}_{\mathrm{d}}} \right\|^{2}}\left[C\left(\gamma_{\mathrm{d}}\right)
\!-\!\sqrt{\frac{1}{T} V\left(\gamma_{\mathrm{d}}\right)}~Q^{\!-\!1}(\epsilon)\right]\notag\\
&=\left(1-\frac{T_{\mathrm{tr}}}{T}\right)
\mathbb{E}_{\left\|\mathbf{\bar{h}_{\mathrm{d}}} \right\|^{2}}\left[C\left(\gamma^{\mathrm{d}}_{\mathrm{eff}}\left\|\mathbf{\bar{h}_{\mathrm{d}}} \right\|^{2}\right)\right]\notag\\
&~~~~\!-\!\left(1\!-\!\frac{T_{\mathrm{tr}}}{T}\right)
\mathbb{E}_{\left\|\mathbf{\bar{h}_{\mathrm{d}}} \right\|^{2}}\left[ \sqrt{\frac{1}{T} V\left(\gamma^{\mathrm{d}}_{\mathrm{eff}} \left\|\mathbf{\bar{h}_{\mathrm{d}}} \right\|^{2}\right)}~Q^{\!-\!1}(\epsilon)\right],
\end{align}
where $\mathbf{\bar{h}}_{\mathrm{d}} \triangleq \mathbf{\hat{h}_{\mathrm{d}}}/\sigma_{\mathrm{\hat{h}_{\mathrm{d}}}}$ is the normalized channel estimate, $\sigma_{\mathrm{\hat{h}_{\mathrm{d}}}}$ is the standard deviation of $\mathbf{\hat{h}_{\mathrm{d}}}$, $\mathbf{\bar{h}_{\mathrm{d}}} \sim \mathcal{CN}(0, \mathbf{I}_{N_\mathrm{B}})$, and $\gamma^{\mathrm{d}}_{\mathrm{eff}}=\frac{\rho_{\mathrm{b}} \sigma^2_{\mathrm{\hat{h}_{\mathrm{d}}}}}{\rho_{\mathrm{b}} \sigma^2_{\mathrm{\hat{e}_{\mathrm{d}}}}+1}$ is the effective SNR.

We next derive a closed-form expression for the lower bound on the data rate achieved by the downlink channel training in the following theorem.
\begin{theorem}\label{Theorem 2_Description}
The lower bound on the data rate achieved by the downlink channel training in \eqref{R_d} is derived as
\begin{align}\label{Theorem 2}
R_{\mathrm{d}}=\left(1-\frac{T_{\mathrm{tr}}}{T}\right)
\left[\Phi\left(\gamma^{\mathrm{d}}_{\mathrm{eff}},N_\mathrm{B}\right)
-\Psi\left(\gamma^{\mathrm{d}}_{\mathrm{eff}},N_\mathrm{B},T\right)\right],
\end{align}
where the functions of $\Phi\left(\gamma^{\mathrm{d}}_{\mathrm{eff}},N_\mathrm{B}\right)$ and $\Psi\left(\gamma^{\mathrm{d}}_{\mathrm{eff}},N_\mathrm{B},T\right)$ are given by \eqref{Phi} and \eqref{Psi}, respectively.
\begin{IEEEproof}
The proof is similar to the proof of \emph{\textbf{Theorem}~\ref{Theorem 1_Description}} and thus omitted here due to space limitation.
\end{IEEEproof}
\end{theorem}

\section{Determination of Channel Reciprocity for Uplink Channel Training Outperforming Downlink Channel Training}\label{sec:Channel Reciprocity Requirement}

In this section, we examine the minimum channel reciprocity coefficient (i.e., the minimum value of $\phi$, which is denoted by $\phi^{\ast}$) that enables the uplink channel training to outperform the downlink channel training. To this end, we derive a closed-form expression to approximate $\phi^{\ast}$ in the following proposition, which is channel-independent and can be used to select the better strategy between the uplink and downlink channel training in practice.
\begin{proposition}\label{Proposition 1_Description}
The minimum channel reciprocity coefficient $\phi^{\ast}$ that enables the uplink channel training to outperform the downlink channel training is approximated as
\begin{align}\label{Beta}
\phi^{\ast}=\frac{\left(\rho_{\mathrm{b}}+1\right)\left(\kappa-1\right)}
{\rho_{\mathrm{b}}\left(N_\mathrm{B}+\kappa-1\right)},
\end{align}
where $\kappa=\left(1+\gamma^{\mathrm{d}}_{\mathrm{eff}}N_\mathrm{B}\right)^{\frac{T-T_{\mathrm{tr}}^{\ast}}{T}}$ and $T_{\mathrm{tr}}^{\ast}$ is the optimal value of $T_{\mathrm{tr}}$ that maximizes $R_{\mathrm{d}}$ in the downlink channel training.
\begin{IEEEproof}
According to its definition, $\phi^{\ast}$ can guarantee $R_{\mathrm{u}}= R_{\mathrm{d}}$. Following \eqref{Theorem 1} and \eqref{Theorem 2}, we find that in order to guarantee $R_{\mathrm{u}}= R_{\mathrm{d}}$ we need
\begin{align}\label{Beta_1}
&\Phi\left(\gamma^{\mathrm{u}}_{\mathrm{eff}},N_\mathrm{B}\right)
-\Psi\left(\gamma^{\mathrm{u}}_{\mathrm{eff}},N_\mathrm{B},T\right)\notag\\
&=\left(1-\frac{T_{\mathrm{tr}}^{\ast}}{T}\right)
\left[\Phi\left(\gamma^{\mathrm{d}}_{\mathrm{eff}},N_\mathrm{B}\right)
-\Psi\left(\gamma^{\mathrm{d}}_{\mathrm{eff}},N_\mathrm{B},T\right)\right].
\end{align}
We find that it is difficult to obtain an expression for $\phi^{\ast}$ from \eqref{Beta_1} directly. To tackle this, we find that the rate loss (i.e., $\Psi\left(\gamma^{\mathrm{u}}_{\mathrm{eff}},N_\mathrm{B},T\right)$ or $\Psi\left(\gamma^{\mathrm{d}}_{\mathrm{eff}},N_\mathrm{B},T\right)$) is negligible comparing to the channel capacity. Then, we approximate \eqref{Beta_1} as
\begin{align}\label{Beta_2}
\Phi\left(\gamma^{\mathrm{u}}_{\mathrm{eff}},N_\mathrm{B}\right)
&=\left(1-\frac{T_{\mathrm{tr}}^{\ast}}{T}\right)
\Phi\left(\gamma^{\mathrm{d}}_{\mathrm{eff}},N_\mathrm{B}\right).
\end{align}
Although a closed-form solution for $\phi^{\ast}$ is still mathematically intractable, we present an accurate approximation based on the Jensen's inequality. That is, when $\chi$ is a concave function, we have $\mathbb {E} \left[\chi (x)\right] \leq \chi \left(\mathbb {E} [x]\right)$. We then approximate \eqref{Beta_2} as
\begin{align}
\log_{2}\left(1\!+\! \gamma^{\mathrm{u}}_{\mathrm{eff}} \mathbb{E}\left[\left\|\mathbf{\bar{h}_{\mathrm{u}}} \right\|^{2} \right] \right)
&\!\!=\!\!\left(1\!-\!\frac{T_{\mathrm{tr}}^{\ast}}{T}\right) \log \left(1\!+\!  \gamma^{\mathrm{d}}_{\mathrm{eff}} \mathbb{E}\left[\left\|\mathbf{\bar{h}_{\mathrm{d}}} \right\|^{2} \right] \right),\notag
\end{align}
which leads to
\begin{align}\label{Beta_4}
\log_{2}\left(1+\gamma^{\mathrm{u}}_{\mathrm{eff}} N_\mathrm{B}\right)&\overset{(a)}{=} \left(1-\frac{T_{\mathrm{tr}}^{\ast}}{T}\right)\log_{2}\left(1+\gamma^{\mathrm{d}}_{\mathrm{eff}} N_\mathrm{B}\right),
\end{align}
where $(a)$ holds since $\mathbb{E}\left[\left\|\mathbf{\bar{h}_{\mathrm{u}}} \right\|^{2}\right]=\mathbb{E}\left[\left\|\mathbf{\bar{h}_{\mathrm{d}}} \right\|^{2}\right]=N_\mathrm{B}$.

After performing some algebraic manipulations, we reach the desired result given in \eqref{Beta} following \eqref{Beta_4}, which completes the proof of \emph{\textbf{Proposition}~\ref{Proposition 1_Description}}.
\end{IEEEproof}
\end{proposition}

\section{Numerical Results and Discussions}\label{sec:numerical results}

In this section, we present numerical results to examine the effectiveness of our analysis and solution, including our newly derived closed-form expressions for the lower bounds on the achievable data rates and the approximation of the minimum channel reciprocity coefficient $\phi^{\ast}$.

\begin{figure}[t!]
\centerline{\includegraphics[width=0.85\columnwidth,height=2.4in]{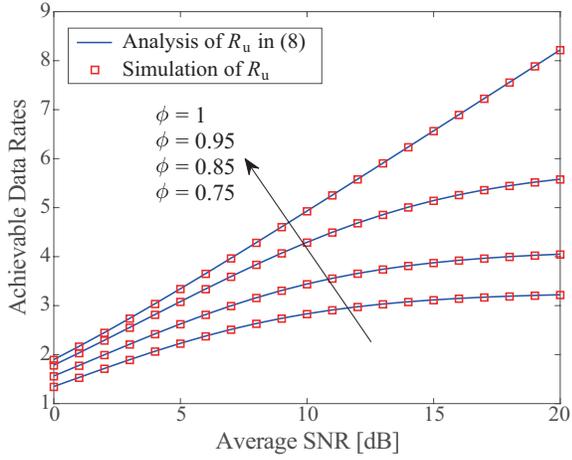}}
\caption{The lower bound on the data rate achieved by the uplink channel training versus the average SNR $\rho_{\mathrm{b}}$ for different values of $\phi$ with $N_{\textrm{B}}=5$, $\epsilon=10^{-9}$, and $T=200$.}\label{Fig:1}\vspace{-2mm}
\end{figure}

In Fig.~\ref{Fig:1}, we demonstrate the accuracy of our newly derived closed-form expression for the lower bound on the data rate achieved by the uplink channel training.
The simulated and theoretical results are obtained from (\ref{R_u}) and (\ref{Theorem 1}), respectively. In Fig.~\ref{Fig:1}, we first observe that the theoretical curves precisely match the simulated ones, which confirms the correctness of (\ref{Theorem 1}) in \emph{\textbf{Theorem}~\ref{Theorem 1_Description}}. Moreover, as expected, in this figure we observe that the data rate significantly increases with $\phi$. Furthermore, we observe that the data rate approaches a constant (but not infinity) as $\rho_b \rightarrow \infty$ when $\phi < 1$, while this data rate increases to infinity as $\rho_b \rightarrow \infty$ when $\phi = 1$. This is due to the fact that the lower bound on the data rate is a linear function of $\rho_{\mathrm{b}}$ when $\phi=1$, since the effective SNR $\gamma^{\mathrm{u}}_{\mathrm{eff}}$ becomes $\rho_{\mathrm{b}}$ as $\rho_b \rightarrow \infty$ for $\phi=1$. Differently, the lower bound is limited by the interference caused by the imperfect channel reciprocity, since $\gamma^{\mathrm{u}}_{\mathrm{eff}}$ becomes $\phi/(1-\phi)$ as $\rho_b \rightarrow \infty$ for $\phi<1$.

\begin{figure}[t!]
\centerline{\includegraphics[width=0.85\columnwidth,height=2.4in]{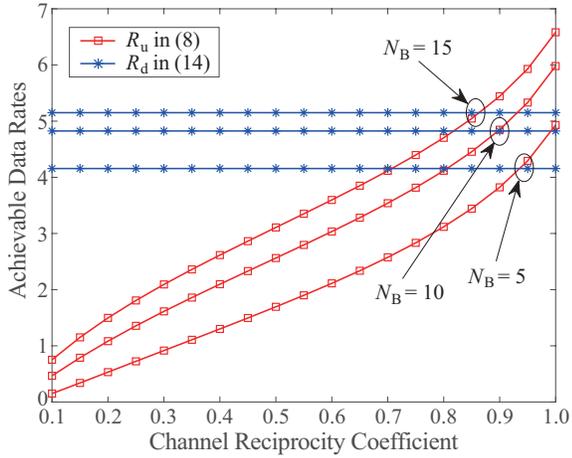}}
\caption{The lower bounds on the data rates achieved by the uplink and downlink channel training strategies versus the channel reciprocity coefficient $\phi$ for different value of $N_\mathrm{B}$ with $\rho_{\mathrm{b}}=10$ dB, $\epsilon=10^{-9}$, and $T=200$.}\label{Fig:2}\vspace{-2mm}
\end{figure}

Fig.~\ref{Fig:2} plots the lower bounds on the data rates achieved by the uplink and downlink channel training versus $\phi$ for different values of $N_{\mathrm{B}}$. The curves for the uplink and downlink channel training are obtained from \eqref{Theorem 1} and \eqref{Theorem 2}, respectively. In this figure, we first observe that the date rate achieved by the uplink channel training increases with $\phi$, which meets our expectation. Importantly, the date rate achieved by the uplink channel training becomes higher than that achieved by the downlink channel training when $\phi$ is greater than a specific value, which is $\phi^{\ast}$.
We also observe that $\phi^{\ast}$ decreases as the number of transmit antennas $N_{\mathrm{B}}$ increases. This is due to the fact that when $N_{\mathrm{B}}$ increases, more time slots need to be used to conduct downlink channel training, while the number of time slots used for the uplink channel training does not change (since $N_{\mathrm{B}}$ is the number of receive antennas in the uplink).

\begin{figure}[t!]
\centerline{\includegraphics[width=0.85\columnwidth,height=2.4in]{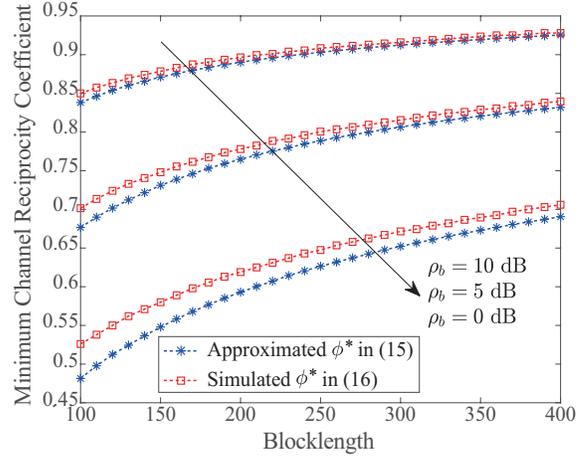}}
\caption{The minimum channel reciprocity coefficient $\phi^{\ast}$ versus the blocklength $T$ for different values of $\rho_{\mathrm{b}}$ with $N_\mathrm{B}=10$ and $\epsilon=10^{-5}$.}\label{Fig:3}\vspace{-2mm}
\end{figure}

In Fig.~\ref{Fig:3}, we examine the accuracy of our approximation of the minimum channel reciprocity coefficient $\phi^{\ast}$, for which the uplink channel training outperforms the downlink channel training. To this end, we compare the simulated $\phi^{\ast}$ obtained based on \eqref{Beta_1} and the approximated $\phi^{\ast}$ obtained from \eqref{Beta}. In Fig.~\ref{Fig:3}, we first observe that the approximated curves are very close to the simulated ones. Also, we observe that the approximation accuracy improves when $\rho_{\mathrm{b}}$ increases, which is due to the fact that the rate loss caused by the finite blocklength becomes negligible when $\rho_{\mathrm{b}}$ tends to be large and this rate loss is not considered in our approximation. In this figure, we further observe that $\phi^{\ast}$ increases with $T$ and $\rho_{\mathrm{b}}$. This is due to the fact that the number of time slots (at least $N_{\mathrm{B}}-1$) or the power saved by the uplink channel training relative to the downlink channel training becomes less significant when $T$ or $\rho_{\mathrm{b}}$ increases, respectively.

\section{Conclusion}\label{sec:conclusion}

In this work, we first fully examined the performance of uplink and downlink channel training strategies. In doing so, we derived closed-form expressions for the lower bounds on the data rates achieved by these two strategies, in which the impact of finite blocklength and channel dispersion was considered. Aided by these expressions, we analytically determined an expression to approximate the minimum channel reciprocity coefficient which enables the uplink channel training to achieve a higher data rate than the downlink channel training. Our examination demonstrated that this minimum channel reciprocity coefficient decreases as the blocklength decreases or the number of transmit antennas increases, revealing the benefits of the uplink channel training in SPC with multiple transmit antennas.

\appendices
\section{Proof of Theorem~\ref{Theorem 1_Description}}\label{Theorem 1 Proof}

In order to prove \emph{\textbf{Theorem}~\ref{Theorem 1_Description}}, we have to solve two integrals in the following:
\begin{align}\label{Phi-Definition}
\Phi\left(\gamma^{\mathrm{u}}_{\mathrm{eff}},N_\mathrm{B}\right)
&=\mathbb{E}_{\left\|\mathbf{\bar{h}_{\mathrm{u}}} \right\|^{2}} \left[C\left(\gamma^{\mathrm{u}}_{\mathrm{eff}}\left\|\mathbf{\bar{h}_{\mathrm{u}}} \right\|^{2}\right)\right]\notag\\
&=\int_{0}^{\infty}\log_{2}(1+\gamma^{\mathrm{u}}_{\mathrm{eff}}x)f_{X}(x)dx
\end{align}
and
\begin{align}\label{Psi-Definition}
\hspace{-2mm}\Psi\left(\gamma^{\mathrm{u}}_{\mathrm{eff}},N_\mathrm{B},T\right)
&=\mathbb{E}_{\left\|\mathbf{\bar{h}_{\mathrm{u}}} \right\|^{2}}\left[\sqrt{\frac{1}{T}}Q^{-1}(\epsilon) \sqrt{V\left(\gamma^{\mathrm{u}}_{\mathrm{eff}} \left\|\mathbf{\bar{h}_{\mathrm{u}}} \right\|^{2}\right)}\right]\notag\\
&=\int_{0}^{\infty} \sqrt{\frac{1}{T}}Q^{-1}(\epsilon)\sqrt{V\left(\gamma^{\mathrm{u}}_{\mathrm{eff}} x\right)} f_{X}(x)dx,
\end{align}
where $f_{X}(x)=x^{N_\mathrm{B}-1}e^{-x}/{\Gamma(N_\mathrm{B})}$ is the probability density function (pdf) of $\|\mathbf{\bar{h}_{\mathrm{u}}} \|^{2}$, since $\mathbf{\bar{h}_{\mathrm{u}}}\sim\mathcal{CN}\left(0,\mathbf{I}_{N_\mathrm{B}}\right)$.

We first tackle the integral in \eqref{Phi-Definition}.
Substituting $f_{X}(x)$ into (\ref{Phi-Definition}), setting $y=1+\gamma^{\mathrm{u}}_{\mathrm{eff}}x$, and using the binomial expansion given in \cite[Eq. (1.111)]{Gradshteyn2007}, we rewrite $\Phi\left(\gamma^{\mathrm{u}}_{\mathrm{eff}},N_\mathrm{B}\right)$ as
\begin{align}\label{Phi-Prove-1}
\Phi(\gamma^{\mathrm{u}}_{\mathrm{eff}},N_\mathrm{B})
&=\frac{e^{\frac{1}{\gamma^{\mathrm{u}}_{\mathrm{eff}}}}\left(\gamma^{\mathrm{u}}_{\mathrm{eff}}\right)^{-N_\mathrm{B}}}{\Gamma(N_\mathrm{B})~\ln{2}} \int_{1}^{\infty}\left(y-1\right)^{N_\mathrm{B}-1}\ln(y) e^{-\frac{y}{\gamma^{\mathrm{u}}_{\mathrm{eff}}}}dy\notag\\
&=\frac{e^{\frac{1}{\gamma^{\mathrm{u}}_{\mathrm{eff}}}}\left(\gamma^{\mathrm{u}}_{\mathrm{eff}}\right)^{-N_\mathrm{B}}}{\Gamma(N_\mathrm{B})~\ln{2}}
\sum_{i=0}^{N_\mathrm{B}-1} \binom{N_\mathrm{B}-1}{i} (-1)^{N_\mathrm{B}-1-i}\notag\\
&~~~\times\int_{1}^{\infty} y^i \ln(y) e^{-\frac{y}{\gamma^{\mathrm{u}}_{\mathrm{eff}}}} dy.
\end{align}
Using the Meijer's G-Function~\cite[Eq. (9.301)]{Gradshteyn2007}, we obtain
\begin{align}\label{Identity-MeijerG}
\int_{1}^{\infty} y^i \ln(y) e^{-\frac{y}{\gamma^{\mathrm{u}}_{\mathrm{eff}}}} dy
\!&=\!\textbf{G}_{2,3}^{3,0}\left({\begin{array}{c} -i,-i \\ 0,-1-i,-1-i \end{array}} \;\middle|\; \frac{1}{\gamma^{\mathrm{u}}_{\mathrm{eff}}}\right).
\end{align}
Substituting (\ref{Identity-MeijerG}) into (\ref{Phi-Prove-1}) we obtain $\Phi\left(\gamma_{\mathrm{eff}},N_\mathrm{B}\right)$ in (\ref{Phi}).

We now solve the integral in \eqref{Psi-Definition}. Substituting $f_{X}(x)$ into \eqref{Phi-Definition} and setting $t=x/\left(N_\mathrm{B}-1\right)$, $\Psi\left(\gamma^{\mathrm{u}}_{\mathrm{eff}},N_\mathrm{B},T\right)$ in \eqref{Psi-Definition} can be rewritten as
\begin{align}\label{Psi-Prove-1}
&\hspace{-4mm}\Psi\left(\gamma^{\mathrm{u}}_{\mathrm{eff}},N_\mathrm{B},T\right)\notag\\
&=\frac{Q^{-1}(\epsilon)}{\Gamma(N_\mathrm{B})} \sqrt{\frac{1}{T}} \int_{0}^{\infty} \sqrt{V\left(\gamma^{\mathrm{u}}_{\mathrm{eff}}~x\right)}e^{\left(N_\mathrm{B}-1\right)\left(\ln x -\frac{x}{N_\mathrm{B}-1}\right)} dx\notag\\
&=\frac{Q^{-1}(\epsilon)}{\Gamma(N_\mathrm{B})} \sqrt{\frac{1}{T}} (N_\mathrm{B}-1)^{N_\mathrm{B}}\notag\\
&~~~~~\times \int_{0}^{\infty} e^{\left(N_\mathrm{B}-1\right)\left(\ln t -t \right)} \sqrt{V\left(\gamma^{\mathrm{u}}_{\mathrm{eff}}~t\left(N_\mathrm{B}-1\right) \right)}~dt\notag\\
&=\frac{Q^{-1}(\epsilon)}{\Gamma(\eta+1)}\sqrt{\frac{1}{T}}\eta^{\eta+1}\int_{0}^{\infty}e^{\eta{}g(t)} \varphi(t)~dt,
\end{align}
where $g(t)=\ln t -t$, $\eta=N_\mathrm{B}-1$, and $\varphi(t)=\sqrt{V\left(\gamma^{\mathrm{u}}_{\mathrm{eff}}t\eta\right)}$.
Then, we can approximate the integral in \eqref{Psi-Prove-1} via the Laplace method as
\begin{align}\label{Psi-Prove-2}
\int_{0}^{\infty} e^{\eta g(t)} \varphi(t) dt
&\approx e^{\eta g(t_0)} \varphi(t_0) \sqrt{\frac{2\pi}{\eta\left|g^{''}(t_0)\right|}},
\end{align}
where $g^{''}(t)=-1/t^{2}$ and $t_0=1$, which is obtained by solving $g^{'}(t)=1/t-1=0$. Finally, substituting $t_0=1$ and \eqref{Psi-Prove-2} into \eqref{Psi-Prove-1}, we obtain the desired result in \eqref{Psi} by performing some algebra manipulations, which completes the proof.

\vspace{2mm}


\end{document}